\documentclass[namedreferences]{solarphysics}

\usepackage[normalem]{ulem}
\usepackage[hyperref,optionalrh]{spr-sola-addons} 
\usepackage{graphicx}        
\usepackage{color}           
\usepackage{breakurl}        
\usepackage{caption}
\usepackage{subfig}
  
\usepackage{amsmath}          
\usepackage{xfrac}

\newcommand{\deriv}[2]{\frac{{\mathrm d} #1}{{\mathrm d} #2}}

\chardef\us=`\_

\newcommand{\pderiv}[2]{\frac{\partial{#1}}{\partial{#2}}}

\begin{document}

\begin{article}
\begin{opening}

\title{Propagation of Leaky MHD Waves at Discontinuities with Tilted Magnetic Field}

\author[corref, email={evickers1@sheffield.ac.uk}]{\inits{E.}\fnm{E.}~\lnm{Vickers}$^{1}$ }
\author[corref]{\inits{I.}\fnm{I.}~\lnm{Ballai}$^{1}$ }
\author[corref]{\inits{R.}\fnm{R.}~\lnm{Erd\'elyi}$^{1}$ }

\institute{
$^{1}$ Solar Physics and Space Plasma Research Centre (SP$^2$RC), School of Mathematics and Statistics, University of Sheffield, Hounsfield Road, Hicks Building, Sheffield, S3 7RH, UK
}

\runningauthor{E. Vickers \textit{et al.}}
\runningtitle{}

\begin{abstract}
We investigate the characteristics of magneto-acoustic surface waves propagating at a single density interface, in the presence of an inclined magnetic field. For linear wave propagation, dispersion relation is obtained and analytical solutions are derived for small inclination angle. The inclination of the field renders the frequency of waves to be complex, where the imaginary part describes wave attenuation, due to lateral energy leakage.
\end{abstract}
\end{opening}

\section{Introduction}
The problem of wave propagation in solar and space plasmas is one of the most important aspects of plasma dynamics. Waves can carry energy across different layers of the solar atmosphere and can dissipate their energy so they contribute to the process of plasma heating in the upper atmosphere \citep[\textit{e.g.}][\textit{etc.}]{einaudi93, erdelyi07, arregui15}. During their progression, waves also carry the imprint of the environment in which they propagate. Therefore, their seismological study can reveal physical parameters that cannot be measured directly or indirectly, such as magnetic field strengths in the tenuous corona, magnitude of various transport coefficients, intrinsic self-organisation of the plasma, \textit{etc.} \citep[\textit{e.g.}][\textit{etc.}]{antia, gough96, gizon05, andries, demoortel12, mathioudakis, arregui18}.

Often waves are propagating along magnetic field lines, which therefore act as tracers, while the structuring of the magnetic fields guide waves along them. While plasma structuring is often concurrent with the structuring of magnetic fields, there are several examples where waves propagate along discontinuities present in plasmas, where the magnetic field is neither parallel nor perpendicular to these interfaces (\textit{e.g.} sunspot penumbra, solar prominences, solar wind, Earth's magnetosphere, \textit{etc.}). Observational and numerical findings confirm large-scale wave propagation along the transition region, for example EIT and Moreton waves \citep[\textit{e.g.}][and references therein]{moreton60, man99, narukage02, ballai05, warmuth15}. In these regions of the solar atmosphere, the ambient magnetic field is already mainly vertical. Large-scale propagating waves could be generated by large-scale eruptions in the solar corona (\textit{e.g.} coronal mass ejections), or by the continuous buffeting of the transition region by solar spicules in the region underneath the transition region \citep[\textit{e.g.}][]{scullion11}.

Several studies have already dealt with the problem of wave propagation inclined to the magnetic field \citep[\textit{e.g.}][\textit{etc.}]{nakagawa73, zhugzhda84, schwartz84, schunker06, schunker08}, but these studies omitted the guided character of waves along discontinuities. Instead of considering a guided nature to waves, these studies either had no plasma structuring, or looked at the transmission or reflection of the waves by an interface. Our research will focus on the propagation of linear waves along interfaces inclined to the magnetic field. 

Discontinuities in magnetised fluids are characterised by the fact that the properties of the plasma change sharply from one equilibrium state to another. These interfaces should be stable against any perturbation in the system, i.e. there is no fluid transport across such structures, in their equilibrium. This translates to the requirement that, in the frame moving together with the discontinuity, the normal component of any displacement is equal on both sides of the interface. For a linear regime, depending on the physical quantities that have to be continuous across discontinuities, we can distinguish three types of discontinuities. First, contact discontinuities are discontinuities where the magnetic field intersects the interface, for which the kinetic pressure ($p$), the magnetic field (${\bf B}$) and velocity (${\bf v}$) are continuous and only mass density and temperatures are allowed to change. The lifetime of such discontinuities is rather short, as in the absence of any restoring force, any displaced plasma element can move freely, eventually leading to the disruption of the discontinuity. In contrast,  for tangential discontinuities, where the magnetic field does not intersect the interface, there is the less strict condition that the total pressure (kinetic and magnetic) is conserved and that only the normal component of the velocity is continuous. The density, kinetic pressure and tangential component of the magnetic field can all be discontinuous at the interface. Finally, for rotational discontinuities the magnetic field and plasma flows change direction but not magnitude. This kind of discontinuity allows mass flow across the interface, but both the density and normal component of the velocity are constant.

The study of MHD waves along discontinuities has concentrated mainly on tangential discontinuities and the literature of this problem is vast \citep[including, for example][\textit{etc.}]{roberts81, hollweg82, ruderman91, jain91, miles92, hau95, joarder06, joarder09, ballai11, mather16, zsamberger18}. The propagation characteristics of these waves are well established. On the contrary, the nature of waves at contact discontinuities and their properties are not so well understood. 

Interfaces, such as the one investigated in the current paper, may also radiate energy to the plasmas away from the interface. The concept of leaky waves under solar conditonswas already investigated by \textit{e.g.} \cite{wilson81}, \cite{spruit82}, \cite{cally85}, \cite{ruderman06}, however none of these included magnetic fields that were inclined to the interfaces.

The structure of our paper is as follows: in Section 2, we introduce the equilibrium model used throughout the rest of this paper and the mathematical formalism. Section 3 is devoted to the derivation of the general governing equation for the plasmas either side of the interface, which is given for an arbitrary inclination of the magnetic field, by Equation \ref{governing}. In order to make analytical progress, in Section 4 we introduce the small inclination angle approximation of the magnetic field, with respect to the interface and solutions to the governing equations for either side of the interface are found, using a perturbation technique. The dispersion relation corresponding to this approximation is obtained in Section 5 and is given in dimensional variables by Equation \ref{eq:DR} and in dimensionless form by Equation \ref{eq:DR_nondim}. In Section 6, solutions to the dispersion relation are determined, both analytically and numerically and finally, our results are summarised and discussed in Section 7.

\section{Initial Equilibrium}
We aim to investigate the propagation of linear and compressible MHD waves along a contact discontinuity in density and temperature, aligned with $z=0$, with density and temperature constant either side of this discontinuity. A constant magnetic field permeates the plasma, at an angle, $\theta$ to the discontinuity in the $(x,z)$ plane and has the form $\mathbf{B_0}=B_0(\cos\theta, 0, \sin\theta)$. 
In our working model, with a tilted magnetic field, the restoring force will be the tangential component of the Lorentz force, with magnetic tension acting on any displacement transversal to the field and magnetic pressure acting on any displacement that changes the magnetic field strength. The effect of gravity is neglected, and we restrict ourselves to study the dynamics of two-dimensional perturbations in a 3D Cartesian geometry. A schematic representation of the equilibrium configuration is shown in Figure 1.

\begin{figure}
\centering
\includegraphics[scale=0.4]{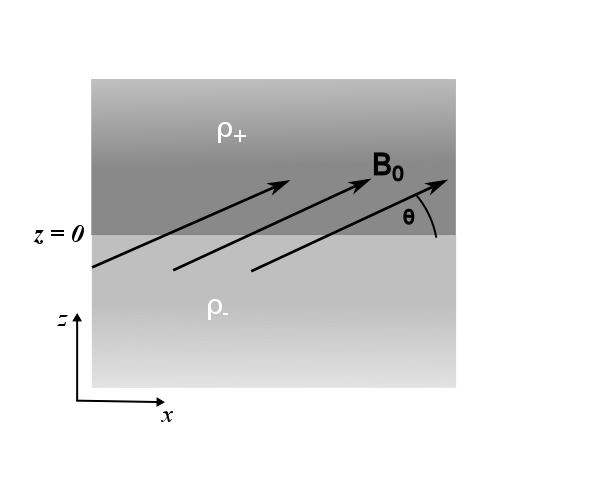}
\caption{The plasma is structured into two infinite regions of constant density and gas pressure, with a sharp interface at $z = 0$. A constant magnetic field crosses the interface and is inclined at an angle $\theta$ to the interface.}
\end{figure}

The dynamics of waves can be described within the framework of ideal magnetohydrodynamics (MHD). We assume that the perturbations of physical quantities are small compared to their equilibrium values, therefore we can use the linearised version of the MHD system of equations. All physical quantities (except velocity) will be written as a sum of their constant equilibrium values and their perturbations, so that the density, pressure, magnetic field and velocity are given by $\rho_{0} + \rho$, $p_{0}+p$, $\mathbf{B}_{0}+\mathbf{b}$ and $\mathbf{v}$, respectively.

The system of ideal linearised MHD equations for a homogeneous equilibrium, therefore, can be written as
\begin{equation}\pderiv{\rho}{t}+\rho_{0}\nabla\cdot \mathbf{v}=0,\label{eq:2.1}\end{equation}
\begin{equation}
\pderiv{p}{t}=c_{0}^{2}\pderiv{\rho}{t},
\label{eq:2.2}
\end{equation}
\begin{equation}
\rho_{0}\pderiv{\mathbf{v}}{t}= -\nabla p - \frac{1}{\mu} \nabla (\mathbf{B}_{0}\cdot \mathbf{b})+\frac{1}{\mu} (\mathbf{B}_{0} \cdot \nabla)\mathbf{b},
\label{eq:mom}
\end{equation}
where $c_0=\left(\gamma p_0/\rho_0\right)^{1/2}$ is the adiabatic sound speed, $\gamma$ is the adiabatic index, and $\mu$ is the magnetic permeability of free space.
In addition, the connection between the magnetic field and the plasma fluid is ensured by the linearised, ideal induction equation,
\begin{equation}
\frac{\partial {\bf b}}{\partial t}=\nabla \times({\bf v}\times {\bf B_0}).
\label{eq:2.5}
\end{equation}
Finally, the system of equations has to be supplemented with the solenoidal condition, $\nabla\cdot {\bf b}=0$, as well as the ideal gas law. These equations will \textbf{be} used in the next section to derive the dispersion relation for waves propagating along the interface.

\section{Governing Equation}

In general the propagation of waves can be studied with the help of their dispersion relation, i.e. the relation between the frequency of waves and the wavenumber, in terms of characteristic speeds and quantities specific to the medium in which they propagate. First, general solutions are found for the plasma regions either side of the interface, the the dispersion relation may be obtained after matching the solutions in the two regions, at the interface $z=0$.

The mathematical procedure we employ in the present study would classify our task as an eigenvalue problem. However, when comparing to an initial value problem investigation, there are differences between a standard eigenvalue problem and the problem concerning leaky waves. As pointed out by \cite{ruderman06}, while standard eigenvalue solutions correspond to the asymptotic behaviour of the time-dependent solutions, the leaky mode solutions are instead intermediate asymptotics, where the time-scale is greater than the period of the wave, but less than the attenuation time.

Let us begin by writing the momentum equation in terms of the inclination angle of the equilibrium magnetic field, $\theta$, as
\begin{equation}
\rho_{0}\pderiv{\mathbf{v}}{t}=-\nabla p-\frac{1}{\mu}B_{0}\nabla \left(b_{x}\cos\theta  + b_{z}\sin\theta \right)+\frac{1}{\mu}B_{0}\left(\cos\theta\pderiv{}{x}+\sin\theta\pderiv{}{z}\right)\mathbf{b},
\end{equation}
where $B_{x}=B_{0}\cos \theta$, $B_{z}=B_{0}\sin\theta$. After differentiating the $x$ and $z$ components of this equation with respect to time and using the energy and induction equations, we obtain two equations for the horizontal and vertical velocity perturbations,
\[
\pderiv{^{2}v_{x}}{t^{2}}=(v_{A}^{2}\sin^{2}\theta+c_{0}^{2})\pderiv{^{2}v_{x}}{x^{2}}+v_{A}^{2}\sin^{2}\theta\pderiv{^{2}v_{x}}{z^{2}}\]
\[ -v_{A}^{2}\sin\theta\cos\theta\pderiv{^{2}v_{z}}{x^{2}}+c_{0}^{2}\pderiv{^{2}v_{z}}{x\partial z}-v_{A}^{2}\sin\theta\cos\theta\pderiv{^{2}v_{z}}{z^{2}},\]
\[
\pderiv{^{2}v_{z}}{t^{2}}=-v_{A}^{2}\sin\theta\cos\theta\pderiv{^{2}v_{x}}{x^{2}}+c_{0}^{2}\pderiv{^{2}v_{x}}{x\partial z}-v_{A}^{2}\sin\theta\cos\theta\pderiv{^{2}v_{x}}{z^{2}}\]
\begin{equation}
 +v_{A}^{2}\cos^{2}\theta\pderiv{^{2}v_{z}}{x^{2}}+(v_{A}^{2}\cos^{2}\theta+c_{0}^{2})\pderiv{^{2}v_{z}}{z^{2}}, \nonumber
\end{equation}
where the Alfv\'en speed is given by $v_{A}=B_{0}/(\mu\rho_{0})^{1/2}$.

All of the perturbed quantities will oscillate with frequency, $\omega$, and real wave-number, $k$. Therefore, we take any perturbations to be of the form $f={\hat f}(z)\exp[i(kx-\omega t)]$, where ${\hat f}$ is the amplitude of perturbations that depend on $z$. Next, we Fourier analyse the above system of equations to arrive at the system of coupled differential equations, 
\[
[\omega^{2}-k^{2}(v_{A}^{2}\sin^{2}\theta+c_{0}^{2})]\hat{v}_{x}+v_{A}^{2}\sin^{2}\theta\deriv{^{2}\hat{v}_{x}}{z^{2}} +k^{2}v_{A}^{2}\sin\theta\cos\theta\hat{v}_{z}+
\]
\begin{equation}
+ikc_{0}^{2}\deriv{\hat{v}_{z}}{z}-v_{A}^{2}\sin\theta\cos\theta\deriv{^{2}\hat{v}_{z}}{z^{2}}=0,
\label{eq:2.10}
\end{equation}
\[
k^{2}v_{A}^{2}\sin\theta\cos\theta\hat{v}_{x} +ikc_{0}^{2}\deriv{\hat{v}_{x}}{z} -v_{A}^{2}\sin\theta\cos\theta\deriv{^{2}\hat{v}_{x}}{z^{2}} +
\]
\begin{equation}
(\omega^{2}-k^{2}v_{A}^{2}\cos^{2}\theta)\hat{v}_{z} +(v_{A}^{2}\cos^{2}\theta+c_{0}^{2})\deriv{^{2}\hat{v}_{z}}{z^{2}}=0.
\label{eq:2.11}
\end{equation}

Since all coefficients in Equations 6 and 7 are constants, we can eliminate the component of velocity parallel to the interface, $\hat{v}_x$ to obtain a single fourth-order differential equation, for the component of velocity perpendicular to the interface, $\hat{v}_z$, 
\[
c_{0}^{2}v_{A}^2\sin^2\theta\deriv{^{4}\hat{v}_{z}}{z^{4}} 
+ 2ikc_{0}^{2}v_{A}^{2}\cos\theta\sin\theta \deriv{^{3}\hat{v}_{z}}{z^{3}} 
+ [\omega^{2}(c_{0}^{2}+v_{A}^{2})-k^{2}c_{0}^{2}v_{A}^{2}]\deriv{^{2}\hat{v}_{z}}{z^{2}} \]
 \begin{equation}- 2ik^{3}c_{0}^{2}v_{A}^{2}\cos\theta\sin\theta\deriv{\hat{v}_{z}}{z} 
 + [\omega^{4}-\omega^{2}k^{2}(c_{0}^{2}+v_{A}^{2})+k^{4}c_{0}^{2}v_{A}^{2}\cos^{2}\theta]\hat{v}_{z}=0.
 \label{governing}
\end{equation}
If the inclination of the magnetic field is omitted (i.e. $\theta=0$), we recover the governing equation for compressional waves derived by \cite{roberts81}. Since we are looking for wave-like behaviour, the solution of the above equation will be of the form $\hat{v}_z \sim\exp[\Gamma z]$. Given that the governing equation (Equation \ref{governing}) is a fourth-order differential equation, the general solution will be of the form 
\[\hat{v}_{z}= C_{1}e^{\Gamma_{1} z} + C_{2}e^{\Gamma_{2} z} + C_{3}e^{\Gamma_{3} z} + C_{4}e^{\Gamma_{4} z},\]
where $C_{i}$ are constants and $\Gamma_{i}$ are the roots of the characteristic equation
\[
c_{0}^{2}v_{A}^{2}\sin^{2}\theta\Gamma^{4} 
+ 2ikc_{0}^{2}v_{A}^{2}\cos\theta\sin\theta \Gamma^{3} 
+ [\omega^{2}(c_{0}^{2}+v_{A}^{2})-k^{2}c_{0}^{2}v_{A}^{2}]\Gamma^{2}
\]
\begin{equation}
- 2ik^{3}c_{0}^{2}v_{A}^{2}\cos\theta\sin\theta\Gamma 
+ [\omega^{4}-\omega^{2}k^{2}(c_{0}^{2}+v_{A}^{2})+k^{4}c_{0}^{2}v_{A}^{2}\cos^{2}\theta]=0.
\label{eq:2.15}
\end{equation}
The values of $\Gamma$ will be used in Section 4 to determine the dispersion relation for the waves propagating along the interface. We note that a similar equation is derived for the plasma regimes both above and below the interface, with appropriate characteristic speeds.

We will show by contradiction that in order to obtain propagating solutions, $\Gamma$ must be complex. Let us assume that $\Gamma$ is purely real, then the left-hand side of Equation \ref{eq:2.15} is a complex analytical function and may therefore be split into real and imaginary parts as
\begin{equation}
u(\omega,k)+iv(\omega,k)=0,
\label{uv}
\end{equation}
where $u$ and $v$ are the real functions,
\[
u(\omega,k) = c_{0}^{2}v_{A}^2\sin^2\theta\Gamma^{4} + [\omega^{2}(c_{0}^{2}+v_{A}^{2})-k^{2}c_{0}^{2}v_{A}^{2}]\Gamma^{2} \]
\[+ [\omega^{4}-\omega^{2}k^{2}(c_{0}^{2}+v_{A}^{2})+k^{4}c_{0}^{2}v_{A}^2\cos^2\theta], 
\]
\[
v(\omega,k) = \Gamma [2kc_{0}^{2}v_{A}^2\sin\theta\cos\theta \Gamma^{2}-2k^{3}c_{0}^{2}v_{A}^2\cos\theta\sin\theta].
\]
In order for Equation \ref{uv} to be satisfied, we require that \emph{both} functions $u(\omega,k)$ and $v(\omega,k)$, are equal to zero simultaneously. Setting $v(\omega,k)=0$ gives the solutions $\Gamma = 0, \pm k$.  The solutions $\Gamma = 0$ is not a solution to $u(\omega,k)=0$. Substituting $\Gamma=\pm k$ into Equation \ref{governing} gives that $\omega =0$, so this is not a propagating solution. This proves that, in order for the wave to propagate, $\Gamma$ must be a complex quantity. The imaginary component of $\Gamma$ represents an oscillatory component to the variation of $\hat{v}_z$ with respect to the transverse coordinate. This will introduce energy flow into the system, either towards or away from the interface, for certain solutions. Since no energy source is specified, this only makes physical sense if energy flow is away from the interface. These \emph{leaky wave} solutions correspond to the case where the group speed is positive above the interface and negative below and the effect of lateral energy leakage is an attenuation of the waves.

\section{Solving The Governing Equation - A perturbation Technique}
In order to make analytical progress, we assume that the angle between the magnetic field lines and the interface is small and so the inclination induces only a small change to the waves' properties in each homogeneous semi-infinite volume, compared to the case with parallel magnetic field. This allows us to make approximations in $\theta$, letting $\cos\theta \approx 1$ and $\sin\theta\approx\theta$.

The governing equation of the quantity $\Gamma$, for small values of inclination angle, is given by
\begin{equation}
c_0^2v_A^2\theta^2\Gamma^4+2ikc_0^2v_A^2\theta\Gamma^3+[\omega^2(c_0^2+v_A^2)-k^2c_0^2v_A^2]\Gamma^2-2ik^3c_0^2v_A^2\theta\Gamma+A=0,
\label{eq:GE}
\end{equation}
where 
\[
A=\omega^4-\omega^2k^2(c_0^2+v_A^2)+k^4c_0^2v_A^2.
\]
In order to physically account for the transition of perturbed quantities from one side of the interface to the other one, we consider a thin boundary layer (embracing the interface), in which the transition takes place, with width less than $2\theta$.

Despite the small angle, even the first term of Equation \ref{eq:GE} is comparable to the other terms, since it is multiplied by the highest derivative of $\hat{v}_z$, which can be large. Let us now apply the method of dominant balance to find the roots of the fourth order polynomial. First, since some terms are smaller than others, they could not possibly be part of the dominant balance and may be ignored. We are going to seek solutions in the form of an asymptotic power series
\begin{equation}
\Gamma=\Gamma_0+\theta\Gamma_1 +\theta^2\Gamma_2+\dots
\label{eq:Gamma}
\end{equation}
In leading order (i.e. terms proportional to ${\cal O}(1))$, we obtain that
\begin{equation}
[\omega^2(c_0^2+v_A^2)-k^2c_0^2v_A^2]\Gamma_0^2-A=0,
\label{eq:Gamma0}
\end{equation}
which means that 
\begin{equation}
\Gamma_0=\pm m,
\label{eq:m0}
\end{equation}
where 
\[
m=\left[-\frac{(\omega^2-k^2v_A^2)(\omega^2-k^2c_0^2)}{(c_0^2+v_A^2)(\omega^2-k^2c_T^2)}\right]^{1/2}.
\]
Interestingly, this quantity coincides with the {\it effective wave-number} determined for magnetoacoustic modes obtained by \cite{roberts81} in the case of tangential discontinuity, if we assume that $m$ is real when $\omega$ is real. However, in the case of leaky modes, $\omega$ is complex and so too is $m$, introducing an oscillatory behaviour in the $z$-direction. A similar expression can be derived for both plasma regions. For simplicity we will introduce the subscripts $+$ and $-$, to refer to equivalent parameters in the plasmas above and below the interface respectively. Hence, in the $z<0$ region we will use $m_-$, whereas in the upper region ($z>0$), we will use $m_+$. Here, the signs of $m_-$ and $m_+$ should be chosen in such a way that the real parts of $m_-$  and $m_+$ are positive. We choose the signs of $\Gamma_0$ above and below the interface, such that these match the solutions for the case of the parallel magnetic field, i.e. physical solutions are given by
\[ \Gamma_0 = \begin{cases} m_-, & \mbox{if } z<0 \\ -m_+, & \mbox{if }z>0. \end{cases} \]

Equation \ref{eq:GE} is a fourth-order polynomial and the remaining two roots must still be found. This can be done by rescaling the problem. For some unknown exponent $Q$ (to be determined), let us set in Equation 11,
\begin{equation}
\Gamma=\theta^Qy,
\label{eq:5}
\end{equation}
where $y$ is bounded and also bounded away from zero as $\theta\to 0$. That is why Equation 11 becomes (after dividing by $c_0^2v_A^2$)
\begin{equation}
\theta^{4Q+2}y^4+2ik\theta^{3Q+1}y^3+\frac{1}{c_T^2}(\omega^2-k^2c_T^2)\theta^{2Q}y^2-2ik^3\theta^{Q+1}y+A'=0,
\label{eq:6}
\end{equation}
where $A'=A/c_0^2v_A^2$ and $c_T=c_0v_A/(c_0^2+v_A^2)^{1/2}$. We will now find the correct value of $Q$ by using the principle of dominant balance, so that the rescaled equation is consistent as $\theta\to 0$, if at least two terms correspond to the same power of $\theta$ (this is called {\it balance}). In addition, the balance is {\it dominant} in the sense that every term not involved in the balance corresponds to a higher power of $\theta$, and therefore must be smaller than the balancing terms.

It can be shown that, when balancing the first three terms, the balance will occur for $Q=-1$ and the balancing terms are proportional to ${\cal O}(\theta^{-2})$ while the other terms  are $\sim {\cal O}(1)$. With this value of $Q$ we obtain that
\begin{equation}
\theta^{-2}y^4+2ik\theta^{-2}y^3+\frac{1}{c_T^2}(\omega^2-k^2c_T^2)\theta^{-2}y^2-2ik^3y+A'=0.
\label{eq:6b}
\end{equation}
Multiplying the above equation by $\mu=\theta^2$ we have
\begin{equation}
y^4+2iky^3+\frac{1}{c_T^2}(\omega^2-k^2c_T^2)y^2-2ik^3\mu y+A'\mu=0.
\label{eq:7}
\end{equation}
Now, we write $y$ also in the form of an asymptotic series in terms of $\mu$ as
\[
y=y_0+y_1\mu+y_2\mu^2+\dots
\]
In the leading order, we obtain 
\begin{equation}
y_0^2(y_0^2+2iky_0+\frac{1}{c_T^2}(\omega^2-k^2c_T^2))=0,
\label{eq:8}
\end{equation}
that has 4 roots for $y_0$, i.e. 
\[
y_{0A}=y_{0B}=0, \quad y_{0(C,D)}=-\frac{i}{c_T}(kc_T\mp \omega)
\]
The $y_0=0$ solutions contradict our assumptions that $y$ is bounded away from zero as $\theta\to 0$ and must be disregarded, so the remaining two solutions will be 
\begin{equation}
y_{0(C,D)}=-\frac{i}{c_T}(kc_T\mp \omega).
\label{eq:9}
\end{equation}
Returning now to the original variables, the four roots of the polynomial in $\Gamma$ are (in the leading order)
\[
\Gamma_A=m + \mathcal{O}(\theta); \quad \Gamma_B=-m+ \mathcal{O}(\theta);\]
\[ \Gamma_C=\theta^{-1}y_{0C}+ \mathcal{O}(\theta); \quad \Gamma_D=\theta^{-1}y_{0D}\mathcal+{O}(\theta).
\]
Higher-order terms in $\theta$ are neglected, as they are small compared to the leading order terms.

Let us briefly discuss the form and sign of the last two roots. The key ingredient in both expressions is $y_{0(C,D)}=-\frac{i}{c_T}(kc_T\mp \omega)$. Since we are dealing with the an inclined interface, as mentioned earlier, we expect that modes will be leaky and the frequency of waves can be written as $\omega=\omega_r+i\omega_i$. Introducing this expression into the form of $y_{0(C,D)}$ we obtain that 
\begin{equation}
y_{0C}=-\frac{\omega_i}{c_T}+\frac{i}{c_T}(\omega_r-kc_T); \quad y_{0D}=\frac{\omega_i}{c_T}-\frac{i}{c_T}(\omega_r+kc_T).
\label{eq:11}
\end{equation}
Since attenuation of waves due to leakage corresponds to $\omega_i<0$, Equation \ref{eq:11} shows that $\Re(y_{0C})>0$, while $\Re(y_{0D})<0$. To ensure that the group speed is positive above the interface and negative below it, the physically acceptable solutions for $y_{0C}$ and $y_{0D}$ must have the corresponding sign. Hence, for the $z<0$ region we are going to use $y_{0D}$, while in the region above the interface we will need to use a similar root as $y_{0C}$ but written for the corresponding characteristic speeds. 

Once again, we use the subscripts $+$ and $-$ to refer to the upper and lower plasma regions and so we write $y_{0D}$, below the interface, as $y_-$ and $y_{0C}$ above the interface as $y_+$.
Therefore the expression of $\hat{v}_z$ in the lower ($z<0$) region is
\begin{equation}
\hat{v}_{z-}=F_-e^{m_- z}+G_-e^{\theta^{-1}y_-z}.
\label{eq:vz>0}
\end{equation}
The corresponding expression for $v_z$,  in the upper ($z>0$) region is
\begin{equation}
\hat{v}_{z+}=F_+e^{-m_+ z}+G_+e^{\theta^{-1}y_+z},
\label{eq:vz<0}
\end{equation}
where the critical speeds for the relevant plasma are used to determine $m_\pm$ and $y_\pm$ is that region.

It is clear that, as $\theta\to 0$, the expressions for $\hat{v}_{z\pm}$ do not approach the variables we would obtain for the tangential discontinuity \citep[for their expressions see, \textit{e.g.}][]{roberts81}, due to the terms proportional to $G_\pm$. As previously mentioned, the amplitude of the transversal component of velocity varies over $z$ in an oscillatory fashion, due to the complex nature of the exponents. In fact the amplitude of the $z$-component of the velocity increases as the inclination angle decreases. Equally, the laterally oscillating leaking wave will have its wavelength increasing proportionally to the inclination angle. 

In a recent paper, \cite{ruderman_vickers18} have shown that, in the case of a tilted magnetic field, the perturbed quantities do not show a simple transition from their values at the contact discontinuity to the tangential discontinuity, as $\theta \to 0$, however their averaged values tend towards those corresponding to the tangential discontinuity. This averaging technique may also be applied to the solutions found here for $v_z$, confirming the above statement.

\section{Dispersion Relation of Waves Along Discontinuities}
Since four constants are involved in the two expressions of $\hat{v}_z$, {\it four} boundary conditions will be needed to find the values of those constants and determine the dispersion relation. As the magnetic field intersects the interface, this is a contact discontinuity and so we require continuity of both components of velocity, $\hat{v}_x$ and $\hat{v}_z$, the kinetic pressure, $\hat{p}$, and both components of the magnetic field, $\hat{b}_x$ and $\hat{b}_z$, respectively. In this equilibrium, continuity of $b_z$ is implied by the continuity of $\mathbf{v}$, so we have the correct number of boundary conditions to find all unknowns.

First, we must find $\hat{v}_x$. Using the components of the momentum equation, we can find that $\hat{v}_{x}$ may also be written in terms of the same exponentials as $\hat{v}_{z}$, i.e.
\[\hat{v}_{x\pm}=f_\pm F_\pm\exp (\mp m_\pm z) + g_\pm G_\pm\exp (y_\pm z),\]
where the expressions for $f_\pm$ and $g_\pm$ are given by
\begin{equation}
f_\pm= \frac{\theta v_{A_\pm}^2(m_\pm^2-k^2)\mp ikc_{0\pm}^2m_\pm}{\omega^2-k^2c_{0\pm}^2}, \quad 
g_\pm= \frac{1}{\theta}\frac{v_{A_\pm}^2y_\pm^2 - \theta kc_{0\pm}^2y_\pm}{(\omega^2-k^2c_{0\pm}^2)}=\frac{1}{\theta}g'_\pm.
\end{equation}

The condition that the vertical component of velocity, $\hat{v}_z$, is continuous across the boundary gives
\begin{equation}
F_-+G_-=F_++G_+.
\end{equation}
Continuity of the parallel component of velocity, $\hat{v}_x$, results in
\begin{equation}
\theta f_-F_-+g'_-G_-=\theta f_+F_++g'_+G_+.
\end{equation}
The condition that the parallel component of the magnetic field, $b_x$, is continuous, gives
\begin{equation}
-\theta m_-F_- + (g'_- -1)y_-G_- = \theta m_+F_+ + (g'_+ -1)y_+G_+,
\end{equation}
to first order of $\theta$. Finally, pressure balance across the interface gives
\[\rho_-c_{0-}^2\left[\theta(ikf_-+m_-)F_- + (ikg'_-+y_-)G_- \right]\]
\begin{equation}
= \rho_+c_{0+}^2\left[\theta(ikf_+-m_+)F_+ + (ikg'_++y_+)G_+ \right].
\end{equation}
Due to the particular choice of discontinuity, $\rho_-/\rho_+ = c_{0+}^2/c_{0-}^2$, so multipliers cancel in this equation.

These conditions may be written together as a matrix equation,
\[
\mathbf{M}
\begin{bmatrix}
F_-\\ G_- \\ F_+\\ G_+
\end{bmatrix}=0,
\]
where,
\[
\mathbf{M} = 
\begin{bmatrix}
1 & 1 &-1 & -1\\
\theta f_- & g'_- & -\theta f_+ & -g'_+ \\
-\theta m_- & (g'_- -1)y_- &  -\theta m_+ & - (g'_+ -1)y_+\\
\theta(ikf_-+m_-) & ikg'_-+y_- &-\theta(ikf_+-m_+) & -(ikg'_++y_+)
\end{bmatrix}.
\]
We seek eigen-mode solutions, by solving the \emph{dispersion relation}, which is given by the equation
\begin{equation}
\det (\mathbf{M})=0.
\label{eq:DR}
\end{equation}

\subsection{Dispersion Relation in Dimensionless Form}
The two key parameters that can affect the characteristics of waves are the ratio between the densities of the two plasma regions ($d=\rho_-/\rho_+$) and the plasma-$\beta$, defined as $\beta = 2c_0^2/\gamma v_{A}^2$. That is why we determine the dispersion relation in terms of these two parameters and we write the equations in their dimensionless form. We introduce the phase speed, relative to the Alfv\'{e}n speed of the lower plasma, as $\tilde{c}_{ph}=c_{ph}/v_{A-}=\omega/kv_{A-}$. Due to the particular form of the magnetic field, we see that $d = c_{0+}^2/c_{0-}^2 = v_{A+}^2/v_{A-}^2$ and hence $\beta_-=\beta_+=\beta$.

The effective wave-numbers in dimensionless form are
\[\frac{m_-^2}{k^2}=\frac{(\gamma\beta/2-\tilde{c}_{ph}^2)(1-\tilde{c}_{ph}^2)}{\gamma\beta/2-(\gamma\beta/2+1)\tilde{c}_{ph}^2},
\quad 
\frac{m_+^2}{k^2}=\frac{(d\gamma\beta/2-\tilde{c}_{ph}^2)(d-\tilde{c}_{ph}^2)}{d^2\gamma\beta/2-d(\gamma\beta/2+1)\tilde{c}_{ph}^2} ,\]
and
\[\frac{y_-}{k} = -i-i\tilde{c}_{ph}\sqrt{\frac{\gamma\beta/2+1}{\gamma\beta/2}}, \quad \frac{y_+}{k} = -i+i\tilde{c}_{ph}\sqrt{\frac{\gamma\beta/2+1}{d\gamma\beta/2}}.\]
The ratios between tangential and transversal components of velocity are given in dimensionless form by
\[f_- = \frac{\theta\left(m_-^2/k^2-1\right)+i\beta m_-/k}{\tilde{c_{ph}}^2-\beta}, \quad 
f_+ = \frac{\theta d\left(m_+^2/k^2-1\right)-id\beta m_+/k}{\tilde{c_{ph}}^2-d\beta},\]
\[g_-' = \frac{y_-^2/k^2-\theta\beta y_-/k}{\tilde{c}_{ph}^2-\beta}, \quad
g_+' = \frac{dy_+^2/k^2-\theta d\beta y_+/k}{\tilde{c}_{ph}^2-d\beta}.\]

In this new, dimensionless, form, the dispersion relation is now given by the equation
\begin{equation}
\det (\mathbf{M'})=0.
\label{eq:DR_nondim}
\end{equation}
where,
\[
\mathbf{M'} = 
\begin{bmatrix}
1 & 1 &-1 & -1\\
\theta f_- & g'_- & -\theta f_+ & -g'_+ \\
-\theta \frac{m_-}{k} & (g'_- -1)\frac{y_-}{k} &  -\theta \frac{m_+}{k} & - (g'_+ -1)\frac{y_+}{k}\\
\theta\left(if_-+\frac{m_-}{k}\right) & ig'_-+\frac{y_-}{k} &-\theta\left(if_+-\frac{m_+}{k}\right) & -\left(ig'_++\frac{y_+}{k}\right)
\end{bmatrix}.
\]

\section{Numerical Solutions and Discussion of Results}
In what follows, we solve the dispersion relation (Equation \ref{eq:DR_nondim}) numerically for varying density ratio and plasma-$\beta$. For simplicity we choose to plot only forward propagating waves. We should note here that in the present configuration, the symmetry between forward and backward propagating waves is broken and this is due to the magnetic force that is perpendicular to the equilibrium magnetic field, affecting the forward and backward propagating waves in a different way. However, the differences between the dimensionless phase speed of forward and backward propagating waves are small due to the chosen approximation of small field inclination. It is very likely that significant differences may be obtained for arbitrary angles, but this would require a full numerical investigation.

\begin{figure*}
\centering
\subfloat[$d = 0.1$]{\label{fig:W_v_b_d=01}\includegraphics[width=01\textwidth]{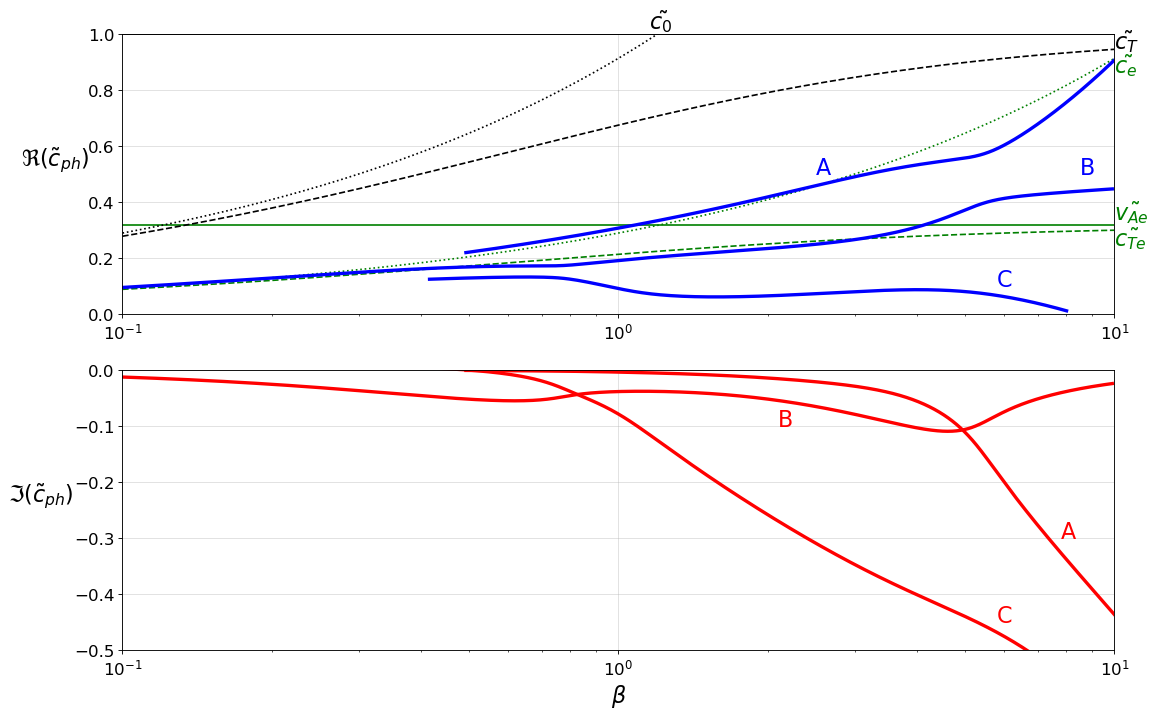}}\\
\vspace{-0.2cm}
\subfloat[$d = 10$]{\label{fig:W_v_b_d=10}
\includegraphics[width=.98\textwidth]{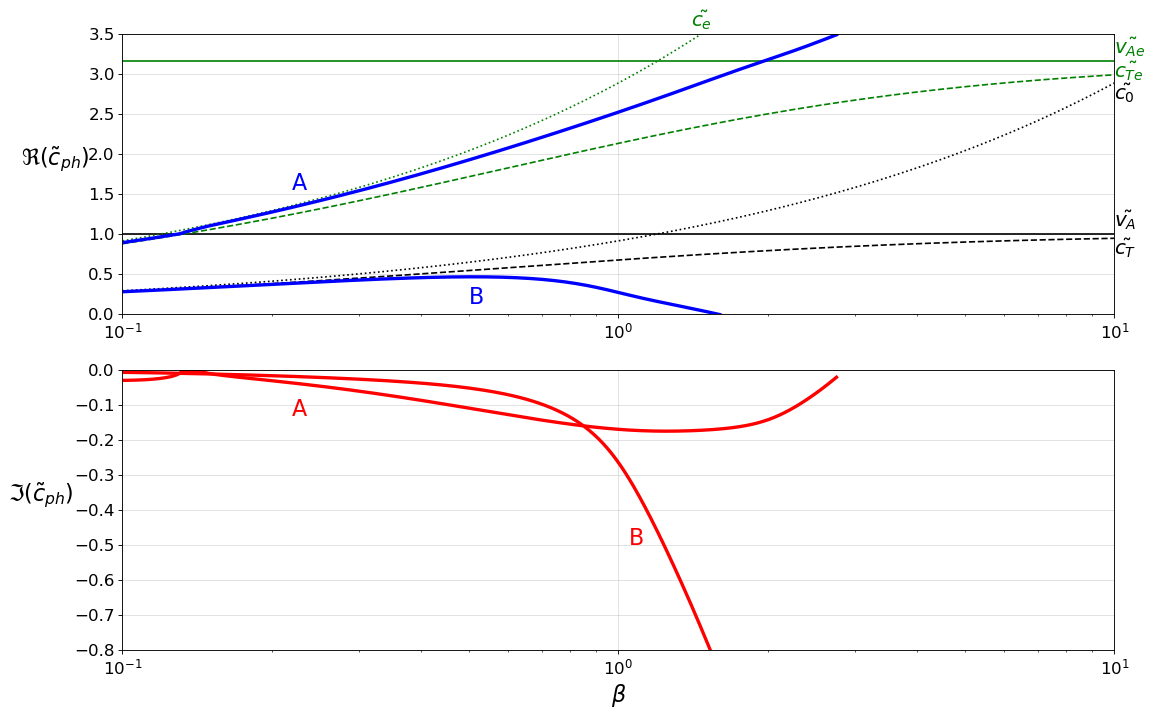}}\\
\caption{The variation of the dimensionless phase-speed of the waves, $\tilde{c}_{ph}$, propagating along the interface in terms of the plasma-$\beta$, for two values of density ratio. The real part is plotted in the \textit{upper panel} and the imaginary part in the \textit{lower panel}. The characteristic speeds are also shown for reference, using \textit{thin lines}: the Alfv\'{e}n (\textit{solid line}), sound (\textit{dotted line}) and cusp speeds (\textit{dashed line}) for the regions above (\textit{green lines}) and below the interface (\textit{black lines}).}
\label{fig:W_v_beta}
\end{figure*}

First, we plot the dispersion curves of forward propagating waves for a fixed density ratio. In Figure \ref{fig:W_v_beta}, we set $d=0.1$ (upper panel) and $d=10$ (lower panel) and allow the plasma-$\beta$ to vary over two orders of magnitude covering the spectrum of possible values in the solar atmosphere plasma. When the plasma above the interface is heavier (here $d=0.1$) the dispersion relation allows the propagation of three modes. Mode (A), is present for larger values of plasma-$\beta$ ($\beta>0.6$), with phase-speed close to the sound speed of the upper plasma, $c_e$. The imaginary part of the solution increases in magnitude for higher values of plasma-$\beta$, so that for $\beta=10$, the period is approximately half of the attenuation time. Since the phase speed is higher than the Alfv\'{e}n speed in the upper region, this mode must be a highly attenuated fast mode. Its propagation speed increases with plasma-$\beta$.

Mode (B) is present for the entire range of plasma-$\beta$ and the phase-speed is, for most of the range, close to the cusp-speed of the plasma in the upper region, $c_{Te}$, increasing to a speed between the two cusp-speeds for high plasma-$\beta$. For $\beta\to0$, we find that mode (B) also tends to zero, meaning that we are dealing with a slow wave. The imaginary part of the solution stays fairly steady, between 0 and -0.1, so these waves show a rather weak attenuation. 

Mode (C), is only present for $\beta>0.5$, since for lower values of plasma-$\beta$, solutions do not satisfy the conditions set for the imaginary part of the frequency (they do not "leak-out"). For $\beta>1$, the phase-speed of this mode decreases, with increasing plasma-$\beta$, while the imaginary part increases in magnitude greatly, so that, for $\beta>1$, the period of the wave is greater than the attenuation time, meaning that the expected life-time of these modes is rather short. This mode should be labelled as a slow mode, as it is slower than either cusp speeds, however, it shows a rather peculiar behaviour. Its phase speed does not increase with plasma-$\beta$, a feature characteristic for slow waves.  Due to their high attenuation, modes (A) and (C) are unlikely to be observable, due to having an attenuation time much shorter than the wave period, especially in the high-beta regime. 

In Figure \ref{fig:W_v_b_d=10}, we show solutions to the dispersion relation for $d=10$ and we can see two modes of propagation. Mode (A) has a phase speed between the sound and cusp speeds in the upper plasma region, though is only present for $\beta<3$, since for higher plasma-$\beta$ the mode is no longer a leaky mode. For low values of plasma-$\beta$, mode (A) tends to zero, so this is clearly a slow mode. In the region corresponding to $\beta<1$, the attenuation of this mode is very small, however, the attenuation rate increases with the value of plasma-$\beta$.  At a value of $\beta=2$ the attenuation of the mode decreases again.  

Mode (B), similar to the  mode (C) in Figure\ref{fig:W_v_b_d=01}, has phase speed close to the lowest value of the cusp-speeds for lower plasma-$\beta$ and its phase speed decreases to zero around $\beta=1$. Again, based on its propagation speed we label this mode as a typical slow mode, however, it has the same peculiar behaviour with respect to plasma-$\beta$.  This mode, again, is highly attenuated, especially for $\beta>1$.

\begin{figure*}
\centering
\subfloat[$\beta = 0.1$]{\label{fig:W_v_d_b=01}\includegraphics[width=1\textwidth]{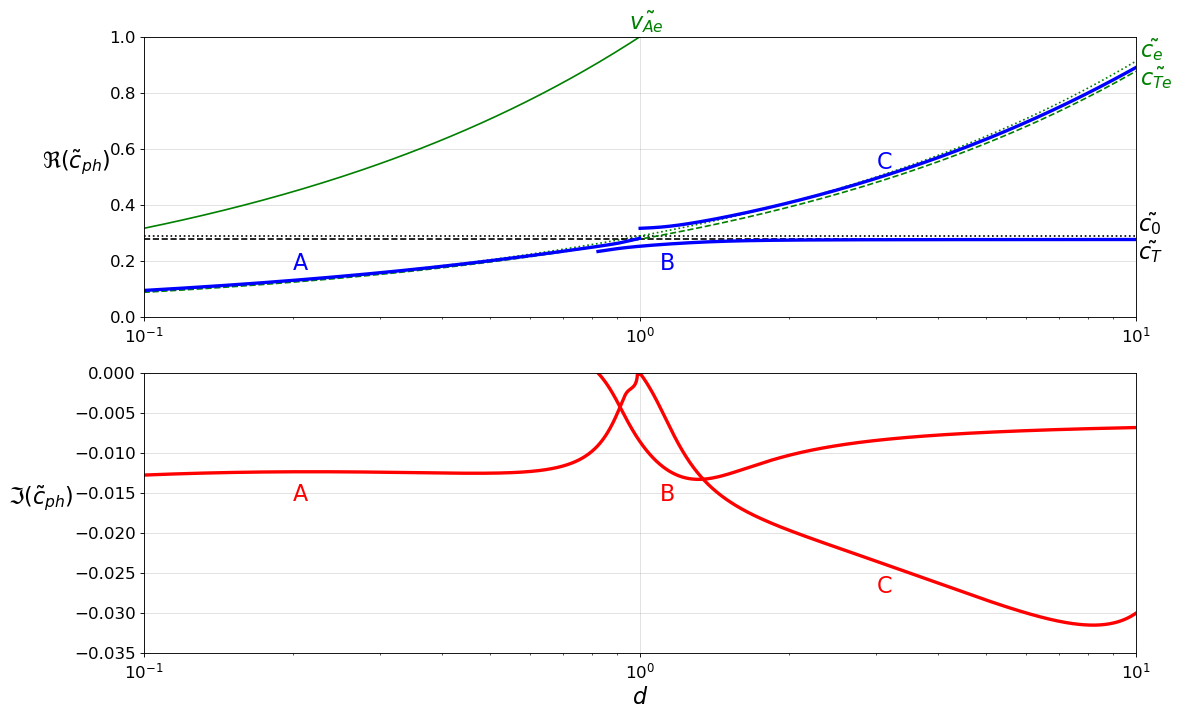}}\\
\vspace{-0.2cm}
\subfloat[$\beta = 10$]{\label{fig:W_v_d_b=10}\includegraphics[width=1\textwidth]{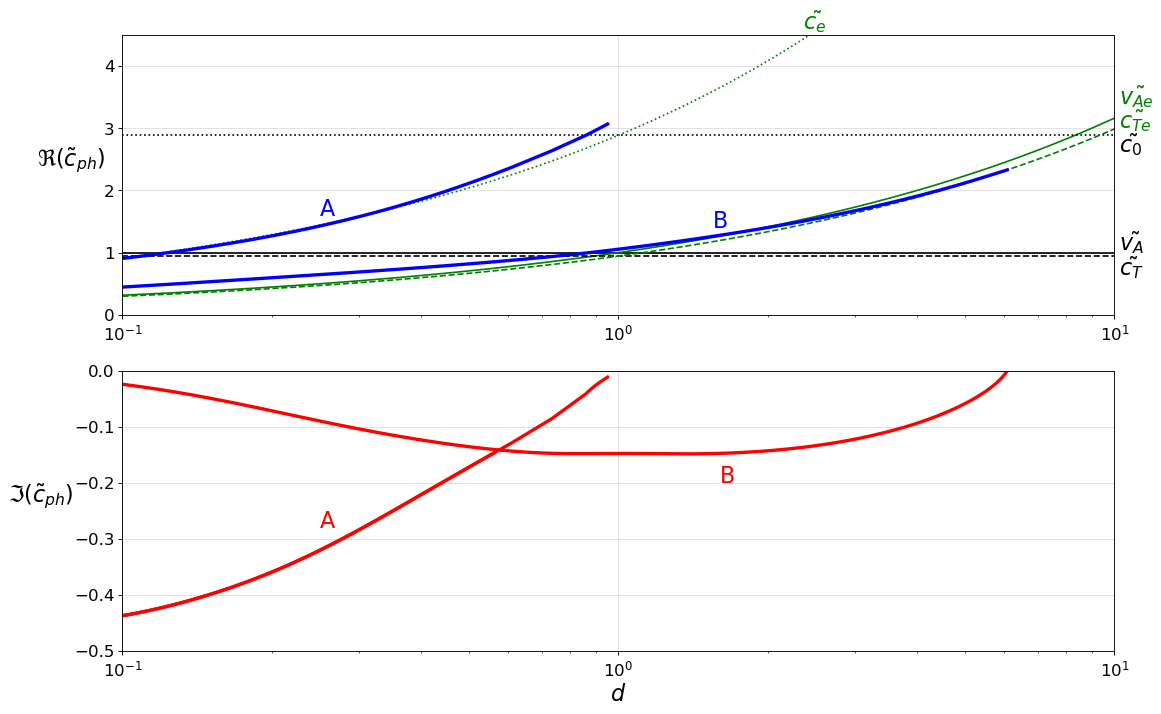}}\\
\caption{The variation of the dimensionless phase-speed of the waves, $\tilde{c}_{ph}$ propagating along the interface in terms of the density ratio, for two values of plasma-$\beta$. The real part is plotted in the \textit{upper panel} and the imaginary part in the \textit{lower panel}. The characteristic speeds are also shown for reference, using \textit{thin lines}: the Alfv\'{e}n (\textit{solid line}), sound (\textit{dotted line}) and cusp speeds (\textit{dashed line}) for the regions above (\textit{green lines}) and below the interface (\textit{black lines}).}
\label{fig:W_v_dens}
\end{figure*}

Figure \ref{fig:W_v_dens} shows the variation of the solutions of the dispersion relation with density ratio, for low and high values of plasma-$\beta$. In Figure \ref{fig:W_v_d_b=01}, we set $\beta=0.1$ (a typical solar upper solar atmospheric condition) and the numerical investigation of the dispersion relation reveals the existence of three modes, all with relatively small imaginary part, meaning these modes are weakly attenuated. Mode (A) has phase speed below the cusp and sound speeds of the plasma in the lower region and a small imaginary part, which decreases in magnitude towards $d=1$, at which point the imaginary part becomes zero. Mode (B) is a physical solution for $d>0.8$ and has phase speed very close to the cusp speed of the lower plasma region. Although the dimensionless phase speed of these two modes are very similar in the region where $d=1$, they have a rather distinctive imaginary part. Mode (C) exists for $d>1$ and has phase-speed close to both the sound an cusp speeds of the upper plasma and this mode shows the largest attenuation among all possible modes. The $d=1$ value corresponds to the situation when the difference between the two regions disappear and there is no interface. This situation was earlier studied by \cite{cally06} and the dispersion relation for MHD waves in this context, for small inclination angle between the wave vector and the magnetic field, may be easily solved to give that the $c_{ph} \approx c_0$ or $v_A$. This agrees with the value of mode (C) at this point and explains the decrease in attenuation of modes (A) and (C), towards $d=1$.  

In Figure \ref{fig:W_v_d_b=10}, we plot the variation of the dimensionless phase speed of waves for a given value of plasma-$\beta$ (here taken to be 10) and we let the density ratio of the two regions vary. This regime of parameters is more relevant to photospheric conditions. We can see, the interface enables the propagation of two surface modes. Comparing these solutions to the ones obtained for low coronal conditions (i.e. $\beta<1$) it is obvious that under photospheric conditions these modes have a much stronger attenuation, the leakage of waves is more accentuated in plasmas with $\beta>1$.  Mode (A) has phase speed very close to the sound speed of the upper plasma region and an imaginary part to the solutions with fairly large amplitude, which decreases towards zero, with increasing density ratio. Given its propagation speed, this mode is a fast MHD mode. This mode does not exist when the plasmas in the upper region becomes heavier. In this case, mode (A) does not satisfy the condition imposed on the imaginary part of its frequency.  Mode (B) shows a rather interesting characteristics as its propagation speed is sub-Alfv\'{e}nic for $d<1$ (the Alfv\'en speed in the lower region is taken as reference), and it becomes super-Alfv\'enic for $d>1$. For the entire domain of its definition, the propagation speed of waves stays close to the sound speed $c_0$. Given the lower phase-speed, as well as a smaller degree of attenuation, we suggest that mode (B) is a slow MHD mode.

\section{Conclusion}
Our study considers the problem of waves propagating in a two-dimensional configuration along an interface in the presence of an inclined magnetic field. For a small-angle approximation, the dispersion relation is derived analytically, by employing the effective wavenumbers and using a method of dominant balance. The effective wavenumbers are shown to be complex, so wave amplitude decays laterally in an oscillatory pattern away from the interface. The complex effective wavenumbers, in turn give rise to complex solutions for the frequency of the waves, which would give amplification or attenuation. However, since no outside energy source exists in this situation, the only physical solutions are leaky waves.

Solutions to the dispersion relation, for varying density ratio and plasma-$\beta$, were found numerically. The imaginary components to these solutions imply attenuation of the wave, which corresponds to energy leaking away from the interface, towards $|z|\to \infty$. We have, hence, shown that, the introduction of magnetic field inclination introduces energy flow to the system, compared to the case with parallel magnetic field and so even a small angle between the interface and the magnetic field produces a qualitative change in the modes which may propagate. Thus, a contact discontinuity in density and temperature in the presence of an inclined constant magnetic field may support the propagation of surface leaky waves. However, quantities averaged over a thin boundary layer do tend towards the solutions for a tangential discontinuity, as the inclination angle tends towards zero, so the contact and tangential discontinuity solutions are qualitatively comparable. 

These results may also have considerable applications to the study of waves in the solar atmosphere. In particular, the penumbra of sunspots have been shown to have highly inclined magnetic field lines and, at high-resolutions, \emph{running penumbral waves} were detected \citep[\textit{e.g.}][]{giovanelli72, zirin72}. Understanding that the outer edge of the sunspot may support leaky waves, more readily than trapped waves, could give a different explanation to any observed damping of these running penumbral waves. While, inside the sunspot, the sharp density variation is horizontal, in the outer edge of the penumbra and the outer canopy, the sharp density gradient is vertical. It has been shown \citep[in \textit{e.g.}][]{jess13} that field inclination (from the horizontal) in the penumbra is $0 - 60^\circ$, but also that the field inclination in the magnetic canopy is $0 - 35^\circ$. Hence, small-angle approximation may have relevance to a much wider range of applications, especially across the solar transition region, where the density variation is sudden. This study may also be of particular use for the investigation of transition region quakes (TRQs), since these waves have large vertical length scales, so the transition region may be viewed as a single interface. When considering TRQs as surface waves propagating along an density discontinuity, the fact that interfaces with inclined fields can only support leaky waves, may help to explain how energy is transferred into the solar corona, through wave leakage.

\section*{Acknowledgements}
E.V. and R.E. acknowledge the support by the Science and Technology Facilities Council UK (grant number ST/M000826/1). I.B. was partially supported by the Leverhulme Trust (IN-2014-016). I.B. was also partly supported by a grant of the Ministry of National Education and Scientific Research, RDI Programme for Space Technology and Advanced Research - STAR, project number 181/20.07.2017. The authors acknowledge the help and advice of M.S. Ruderman.

\textbf{Disclosure of Potential Conflicts of Interest} The authors declare that they have no conflicts of interest.

\bibliographystyle{spr-mp-sola}
\bibliography{contact_bibliography}

\end{article}
\end{document}